\documentclass[10pt]{iopart}

\usepackage{graphicx}				% Use pdf, png, jpg, or eps§ with pdflatex; use eps in DVI mode

\usepackage{hyperref}

\usepackage{float}

\usepackage{iopams}
\usepackage[utf8]{inputenc}
\usepackage{xcolor}
\usepackage{subcaption}
\usepackage[normalem]{ulem}
\usepackage{mathrsfs}
\usepackage{csquotes}
\usepackage{cite}

\usepackage{bbm}
\usepackage{enumitem}

\newcommand\dens{\rho}
\newcommand{\mycomment}[1]{}

\begin{document}

\title{Quantum features of the transport through ion channels in the soft knock-on model}

\author{Mateusz Polakowski and Miłosz Panfil}

\address{Faculty of Physics, University of Warsaw, Pasteura 5, 02-093 Warsaw, Poland}

\date{\today}

\begin{abstract}
    Ion channels are protein structures that facilitate the selective passage of ions across the membrane cells of living organisms. They are known for their high conductance and high selectivity. The precise mechanism between these two seemingly contradicting features is not yet firmly established. One possible candidate is the quantum coherence. In this work we study the quantum model of the soft knock-on conduction using the Lindblad equation taking into account the non-hermiticity of the model. We show that the model exhibits a regime in which high conductance coexists with high coherence. Our findings second the role of quantum effects in the transport properties of the ion channels. 
\end{abstract}

\submitto{\PB}

\maketitle
\ioptwocol

\section{Introduction}
Ion channels form a large family of protein complexes, which role is to regulate flow of different ions across cellular membranes. They are an ubiquitous feature among all excitable cells and can be found in both the largest animals and the smallest bacteria. Since their discovery, they have been the subject of intense study, with dozens of channel families and subfamilies identified along the way. Among them, there are selective potassium channels, which are mainly responsible for reestablishing the resting membrane potential after the upstroke of action potential \cite{She94,Hille92}. One of their most significant features is that in their open state they permeate potassium ions at a near-diffusion rate of around $10^{8} \, \mathrm{ions/s}$, while being 10,000 times more selective in favour of potassium ions over sodium \cite{Hille92}. Crystallization of the ion channels, starting with KcsA potassium channel of the bacteria \textit{Streptomyces lividans}  \cite{Doy98, Sok01, Dut02}, and direct experimental observations of a single channel's action \cite{Cha99, Pos05} allowed us to better understand these proteins on an atomic level. Those and other findings led to the development of mathematical and physical models of permeation \cite{Mor01, Corry00, Corry03, Roux04, Tol06,WAT17}, and enabled for the use of molecular dynamics (MD) simulations \cite{MacKer04, Maf12, Kopf14, Kope18}. However, despite the advances in modelling and experimental probing of ion channels, the problem of seemingly paradoxical simultaneous high conductance and high selectivity has remained unsolved~\cite{Mironenko21}.

Structurally, ion channels are protein complexes inserted into the cell membrane. They are made of multiple subunits, which spatial arrangement forms a pore, through which ions can flow in or out of the cell. The narrowest section of that pore is called selectivity filter (SF), as it is believed that this is the place where the discrimination between ions takes place. In KcsA, the filter is located near the extracellular mouth of the channel. It is made of a sequence of five residues, $\mathrm{T_{75} V_{76} G_{77} Y_{78} G_{79}}$, which are located on the P-loop. Four P-loops of the four subunits delimit a narrow pathway, just 12 {\AA} in length and around 3 {\AA} in diameter, through which the ions must pass in order to reach the extracellular solution. The backbone carbonyl oxygens of Thr 75, Val 76, Gly 77 and Tyr 78 as well as the side-chain hydroxyl of Thr 75 point directly into this pathway, creating four binding sites inside the SF. These binding sites are conventionally labeled S1 through S4, with S1 being at the extracellular site of the SF. Each of the binding sites is able to accommodate one fully dehydrated $\mathrm{K^{+}}$ ion, which is then coordinated by eight negatively charged oxygen atoms, four \enquote{above} the ion and four \enquote{below} it \cite{Doy98,Mor01,Hille92,Arm21}. Although the gating mechanism varies greatly across different types of potassium channels, the amino-acid sequence of the selectivity filter is very similar in all of them \cite{Doy98,Mor01,Bhat18}.

Currently, there are two competing models of ion permeation through potassium ion channels. In the first one, the ions move through the pore separated by water molecules; the binding sites are alternately occupied by ions and water molecules. This model is known as \enquote{water knock-on} or \enquote{soft knock-on} model. It is based on the assumption that due to the electrostatic repulsion between the ions inside the selectivity filter, there can be at most two ions in the SF at any given time. For this reason, the earlier crystallographic experiments showing potassium ions in all four binding sites were interpreted as a superposition of S1-S3 and S2-S4 ion configurations \cite{Doy98,Mor01}. In the competing model, called \enquote{hard knock-on}, the conduction occurs without the presence of water. The ions jump between the binding sites as a result of the short-range ion-ion interaction inside the SF. This model is supported by more recent studies: molecular dynamics simulations \cite{Kopf14,Kope18}, X-ray diffraction measurements \cite{Lan18} and solid-state NMR experiments \cite{Oster19}. Although recent computational and experimental findings seem to favour the \enquote{hard knock-on} model, the debate has not been concluded, especially since the experimental observation of the ion motion through the channel has remained elusive. Hence, the \enquote{soft knock-on} has not been disproved and is still widely studied \cite{Krat16,Sum19,Ryan23,Ryan24}. Moreover, exploring its general properties may be beneficial with regard to other ion channel families, which could employ this conduction mechanism \cite{Ulm13, Liang21, Li23}.

The discovery of long lived quantum coherence in photosynthetic energy transfer demonstrated the importance of quantum phenomena in transport processes of biological systems \cite{Eng07,Lee07,Col10}. This quantum coherence survives despite the dephasing noise originating from fluctuations of an environment. Furthermore, in certain cases, the environmental coupling enhances the efficiency of the system, with its optimal regime being neither completely quantum, nor completely classical \cite{PleHue08, Reb09}.

Since ion channels operate on the sub-nanoscale, both spatially and temporally, they are a natural candidate to look for the quantum effects. Vaziri and Plenio suggested that quantum coherence in SF may play a role in selectivity and conduction process \cite{Vaz10}. They proposed a model in which an ion can hop between the adjacent sites via quantum tunneling or thermal activation. Interplay between quantum coherence and dephasing noise is then essential for conduction properties. De March \textit{et al.} took this idea further by including Coulomb repulsion between the ions inside SF~\cite{Mar18, Mar21}. Summhammer \textit{et al.} \cite{Sum20} found that quantum mechanical MD simulations yield higher conduction rates compared to classical MD. Salari \textit{et al.} claim that the classical coherence between carbonyl groups oscillations is not sufficient to explain high conduction rates and selectivity \cite{Sal15}. In~\cite{Sal17} Salari \textit{et al.} investigated the possibility of quantum interference of potassium ions in neighbouring channels. However, they concluded that the coherence times are too short to play any significant role. 

In this paper we focus on the water-mediated transport method and taking inspiration from the work of Seifi \textit{et al.} \cite{Sei22}, we represent the transport sequence as a three state system, resembling quantum spin-1 system. However, the model is non-hermitian which causes the standard Lindblad approach inapplicable. To overcome problems that it causes, as encountered in~\cite{Sei22}, we use an adjusted Lindblad framework, which guarantees that the density matrix stays positive-definite and of trace $1$ throughout the whole time evolution. We also consider a generalization of the three states model to account for bidirectional transition rates.

Our work is organized as follows. We start with a short description of the Lindblad equation for Hermitian and non-Hermitian Hamiltonians and measures of coherence. We then introduce in details two models of the transport through ion channels. In the following section we present the results and discuss the late-time values of conductance and coherence. The results show that the dynamics of the system has two regimes which differ in the properties of the stationary states. We explain this phenomena using the so-called effective Hamiltonian technique known from the non-Hermitian quantum mechanics. We conclude our work with a summary and further perspectives. 

\section{Lindblad equation and coherence quantifiers}
Biological phenomena occur almost invariably in a relatively hot and complex environment, which is the opposite of the ideal quantum mechanical setting. The system that is of interest to us interacts with the environment around it and this interaction can drastically change its dynamics. Very large, or even infinite, number of degrees of freedom of the environment means that it is impossible to efficiently describe the whole system only by means of Hamiltonian and the von Neumann equation for the density matrix, even if the total Hamiltonian is known to us (which is usually not the case anyway). Instead, one can describe the evolution of the system of interest, called the open quantum system, with the non-unitary master equation. In the Markovian limit the general form of the Lindblad equation is
\begin{equation}
    \frac{d\rho}{dt} = - i \left[H, \rho \right] + D[\rho]  \textrm{,} \label{eq:lindblad}
\end{equation}
where $\dens$ is the density matrix of the open system, $H$ its Hamiltonian and 
\begin{equation}
    D[\rho] = \sum_k \gamma_k \left( L_k \rho L^{\dagger}_k - \frac{1}{2} \left\{ L^{\dagger}_k L_k, \rho \right\} \right) \label{Lindblad_op}
\end{equation}
is the Lindblad superoperator~\cite{BrePet02}. We have assumed $\hbar$ to be one. The first term on the right side of Lindblad equation describes the unitary evolution of the density matrix. The second part describes non-unitary evolution due to interactions with the environment and is responsible for the loss of coherence. The curly brackets denote the anticommutator, the operators $L_k$ are called the Lindblad jump operators and the coefficients $\gamma_k$ are real and non-negative.

When using a non-Hermitian Hamiltonian we need to adjust the description of the evolution of the density matrix. An appropriate framework (hereafter we refer to it as the \enquote{adjusted framework}) for non-Hermitean systems with parity-time symmetry ($\mathcal{PT}$-symmetric~\cite{RevModPhys.96.045002}) was proposed by Brody and Graefe \cite{BroGra12}. The time evolution of a density matrix $\dens$ is then given by
\begin{equation}
   \frac{d\rho}{dt} = - i [\mathscr{H}, \dens ] - \{ \Gamma, \dens \} + 2 \dens  \Tr (\dens \Gamma ) \textrm{,} \label{eq:nonhermitian time evolution}
\end{equation}
where a non-hermitian Hamiltonian $H$ is decomposed into a sum of Hermitian and anti-Hermitian parts
\begin{equation}
    H = \mathscr{H} - i \Gamma \textrm{,} \label{eq:sergi decomp1}
\end{equation}
with two Hermitian operators

\begin{equation}
    \mathscr{H} = \frac{1}{2} \left( H + H^{\dagger} \right), 
    \qquad \Gamma = \frac{i}{2} \left( H - H^{\dagger} \right) \textrm{.}
\end{equation}

The first two terms in the evolution equation~\eref{eq:nonhermitian time evolution} can be easily inferred by inspecting the standard von Neumann equation for the evolution of the density matrix. However, it turns out that with these two terms alone, the trace of a density matrix would not be conserved. Therefore, a third term is needed that ensures the conservation of the trace. This guarantees the retention of probabilistic interpretation of the density matrix and allows calculating statistical averages of operators. If the Hamiltonian $\mathscr{H}$ is already Hermitian, then $\Gamma = 0$ and one recovers the traditional equation of motion.

To obtain the complete formula for the evolution of the density matrix, we add Lindblad operator~\eref{Lindblad_op} on the right-hand side of eq.~\eref{eq:nonhermitian time evolution}, and therefore work within what Zloshchastiev and Sergi call a \enquote{hybrid} formalism \cite{ZloSer13}. The total evolution of the density matrix is governed by the equation
\begin{equation}
    \frac{d\rho}{dt}  = - i [\mathscr{H}, \dens ] - \{ \Gamma, \dens \} + 2 \dens \Tr (\dens \Gamma ) + \mathcal{D} [ \dens ] \label{eq:total nonhermitian evolution} \textrm{,}
\end{equation}
Equation~\eref{eq:total nonhermitian evolution} is the equation that we use to model dynamics of the ion channels. We will specify the Hamiltonian and Lindblad jump operators $L_k$ when discussing the details of the \enquote{soft knock-on} model. Instead now, we turn our attention to measures of coherence.

The framework for quantifying coherence was laid out by Baumgratz \textit{et al.} \cite{Bau14}. For a function $C:~\mathcal{S} (\mathcal{H}) \longrightarrow \mathbb{R}_{+}$, that maps a density matrix into the set of non-negative real numbers, to be a valid coherence measure the following conditions have to be satisfied:
\begin{enumerate}[topsep=1ex, leftmargin=6ex]
    \item[(C1)] $C( \delta) = 0 $ for every $\delta \in \mathcal{I}$,
\end{enumerate}
where $\mathcal{I}$ is the set of completely incoherent states in a given basis, i.e. diagonal density matrices. A stronger condition can be demanded that $C(\dens)$ is non-zero only, and only if $\dens$ contains any coherence. The coherence measure should not increase under the incoherent operations, i.e. operations that cannot create coherence from incoherent states:
\begin{enumerate}[leftmargin=7.5ex]
    \item[(C2a)] $C(\dens) \geq C (\Phi_{\mathrm{ICPTP}} (\dens))$ for all incoherent completely positive trace-preserving maps $\Phi_{\mathrm{ICPTP}}$;
    \item[(C2b)] $C (\dens) \geq \sum_n p_n C(\dens_n)$ with $\dens_n = K_n \dens K^{\dagger}_n / p_n$ and $p_n = \Tr (K_n \dens K^{\dagger}_n)$, for any set of Kraus operators $\{ K_n \}$ satisfying $\sum_n K^{\dagger}_n K_n = \mathbbm{1}$ and $K_n \mathcal{I} K^{\dagger}_n \subseteq \mathcal{I}$.
\end{enumerate}
Finally, state mixing should only decrease coherence, and therefore the coherence measure should be a convex function,
\begin{enumerate}[leftmargin=6ex]
    \item[(C3)] non-increasing under mixing of quantum states (convexity): $\sum_n p_n C(\dens_n) \geq C (\sum_n p_n \dens_n)$ for any set of quantum states $\{ \dens_n \}$ and any $p_n \geq 0$ such that $\sum_n p_n = 1$.
\end{enumerate}
Sometimes, a term \enquote{coherence quantifier} is used for the functions satisfying C1-C3 , while coherence measures satisfy two additional conditions: uniqueness and additivity under tensor products~\cite{Strel17}.

So far several coherence quantifiers and measures have been identified, including ones based on $l_p$-norms, affinity of coherence, robustness of coherence, coherence cost and so on \cite{Strel17}. One canonical measure of coherence is distillable coherence. It describes the optimal number of maximally coherent states $|\Psi_d\rangle$, which can be obtained from a given state $\dens$ in the asymptotic limit. The maximally coherent state $|\Psi_d\rangle$ of the dimension $d$ is defined as
\begin{equation}
    |\Psi_d\rangle = \frac{1}{\sqrt{d}} \sum_{i=1}^d |i\rangle \textrm{,} \label{eq:max coh state}
\end{equation}
where $\{ |i\rangle \}$ is a chosen basis. Its uniqueness comes from the fact that any $d \times d$ state $\dens$ can be obtained from $|\Psi_d\rangle$ by means of incoherent operations. Distillable coherence, also known as the relative entropy coherence introduced by Baumgratz \textit{et al.} \cite{Bau14}, assumes a simple form
\begin{equation}
    C_d (\dens) = S (\Delta \left[ \dens \right]) - S(\dens) \textrm{,} \label{eq:distillable coherence} 
\end{equation}
where $S (\dens) = - \Tr \left( \dens \log_2 \dens \right)$ is the von Neumann entropy, and $\Delta$ is the dephasing operator, which returns only the diagonal part of the density matrix \cite{WinYan16}.
Another popular coherence quantifier is the $l_1$-norm of coherence, given by the sum of absolute values of the off-diagonal elements
\begin{equation}
    C_{l_1} (\dens) = \sum_{i\neq j} |\dens_{ij}| \textrm{.} \label{eq: cl1}
\end{equation}
In this work we will use the distillable coherence which from now we simply denote by $C(\rho)$. However, both of the coherence meassures have been employed to study quantum biology phenomena, including avian magnetoreception \cite{Kom20,Smith22} and transport through ion channels \cite{Mar18,Sei22}.

\section{Model}

\begin{figure}[h]
    \centering
    \begin{subfigure}[b]{0.15\textwidth}
        \includegraphics[width=\textwidth]{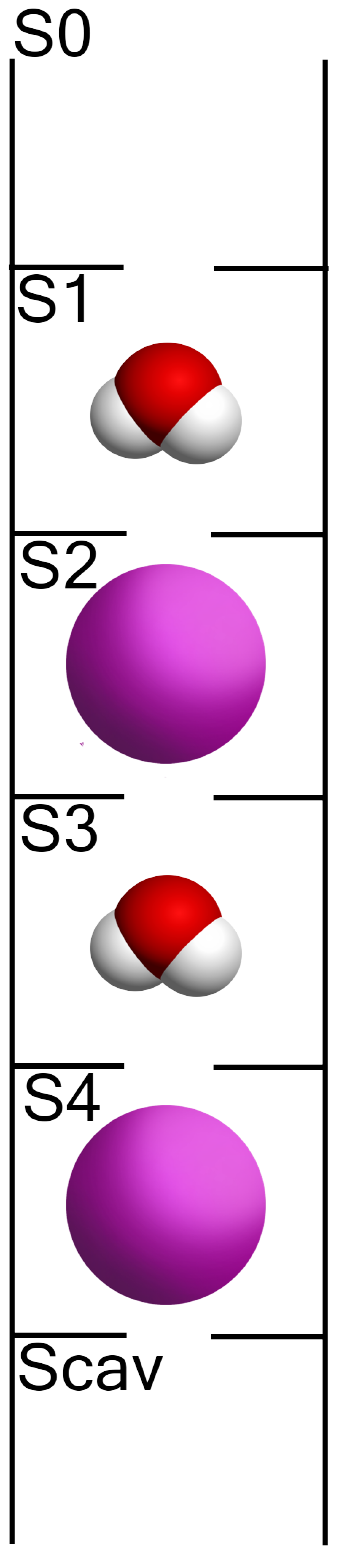}
    \end{subfigure}
    \begin{subfigure}[b]{0.15\textwidth}
        \includegraphics[width=\textwidth]{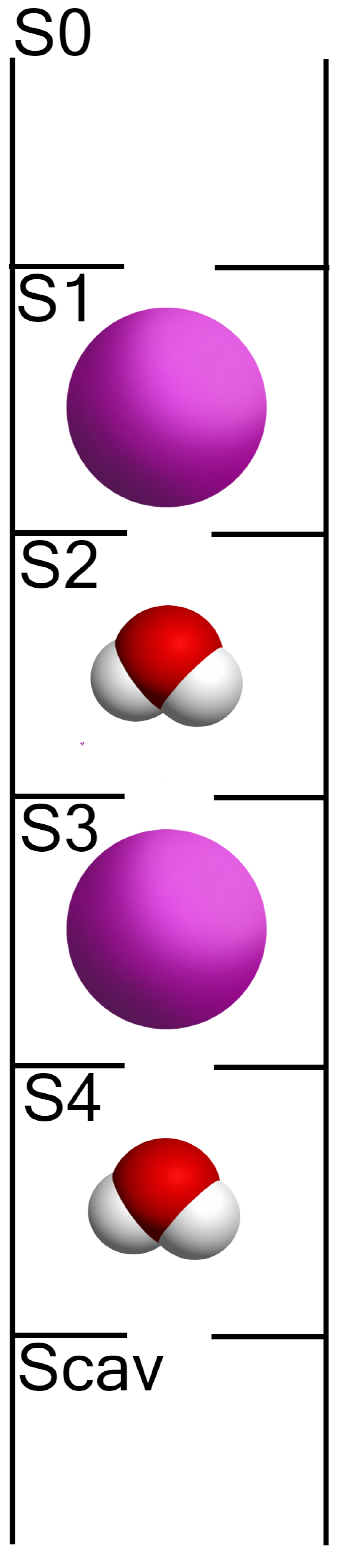}
    \end{subfigure}
        \begin{subfigure}[b]{0.15\textwidth}
        \includegraphics[width=\textwidth]{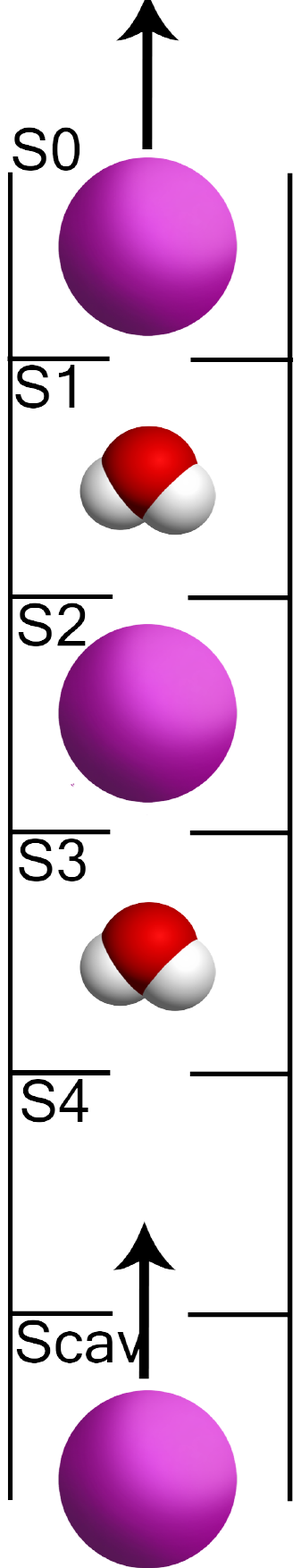}
    \end{subfigure}
    \caption{Schematic illustration of the ion channel states, $|1\rangle$, $|2\rangle$ and $|3\rangle$, from left to right in this order. The purple spheres represent potassium ions.} \label{fig:Seifi 3 states}
 \end{figure}

In the \enquote{soft knock-on} model the transport process of the ion can be conceptualized as proceeding through $3$ states. The three states in question are: potassium ions at the sites two and four and water molecules at the sites one and three, denoted $|1\rangle$; potassium ions at the sites one and three and water molecules at the sites two and four, denoted $|2\rangle$; a potassium ion at the site two and water molecules at sites one and three with an additional potassium ion exiting the ion channel and another one entering the selectivity filter at the site number four, denoted $|3\rangle$. The jump of a potassium ion from the fourth site to the extracellular solution is considered a transition form $|3\rangle$ to $|1\rangle$~\cite{Sei22}. The three configurations are presented graphically in the figure~\ref{fig:Seifi 3 states}.

The Hamiltonian of the system reads
\begin{equation}
    H = \omega_0 S_z + c \left( |2\rangle\langle1|+ |3\rangle\langle2| + |1\rangle\langle3| \right) \textrm{,} \label{eq:Seifi og hamiltonian}
\end{equation}
where $\omega_0$ is a transition frequency, $S_z$ is the z-component of the spin-1 operator and $c$ is the coherent transition rate which, for simplicity, we assume to be the same between different pairs of states. The Hamiltonian can be represented in the matrix form
\begin{equation} 
H = \left(\begin{array}{@{}*{20}{c}@{}}
	\omega_0 & 0 & c\\
	c & 0 & 0 \\ 
	0 & c & -\omega_0
	\end{array}\right) \textrm{.} \label{eq:Ham mat}
\end{equation}
\noindent The Lindblad operator describing the coupling with the environment is of the form
\begin{equation}
    \mathcal{D} [ \dens ] = \gamma \left( S_{-} \dens S_{+} - \frac{1}{2} \left\{ S_{+} S_{-}, \dens \right\} \right) \textrm{,} \label{eq: Seifi og dissipator}
\end{equation}
where $S_{\pm} = \frac{1}{\sqrt{2}} \left( S_x \pm i S_y \right)$ are the spin-1 ladder operators. It models the incoherent transitions from state $|1\rangle$ to $|2\rangle$ and from $|2\rangle$ to $|3\rangle$ which stipulate the transport of the ion through the channel.

The question regarding value range of the parameters used in the model is a difficult one. Vaziri and Plenio provided the first estimation \cite{Vaz10}. They argued that the effective hopping rate parameter $c_{eff} = c^2/\omega_0$ should be of the order of the transfer rate of the channel, which is in the range of $10^6 - 10^8 \, \mathrm{s^{-1}}$. However, they pointed out that this is valid while $c \ll \omega_0$, with $\omega_0$ not larger than $10^{12} \, \mathrm{s^{-1}}$. They themselves set $\omega_0$ to be of an order of magnitude larger than $c$, with $c \sim 10^{9} \, \mathrm{s^{-1}}$. De March \textit{et al.} \cite{Mar18,Mar21} opt for a much larger $c \sim 10^{11} - 10^{13} \, \mathrm{s^{-1}}$, to compensate the electrostatic repulsion and retain the expected effective hopping rate $c_{eff}$. The dephasing rate $\gamma$ varied from a fraction of $c$ up to $100c$. However, note that in the mentioned papers, the authors work with a model based on a tight-binding chain, in which the ions hop independently between the subsequent binding sties. Therefore, transferring these values one-to-one to "soft knock-on" model may not be completely justified. As one may have noticed, there is no general consensus regarding the objective values of the parameters, which therefore are treated rather instrumentally. Although a revised, in-depth discussion would be highly beneficial, currently established values should still provide qualitative results. Hence, we shall take the parameters used by Seifi \textit{et al.} as the starting point but also go beyond them to probe the whole parameter space of the model.

\begin{figure}[ht]
    \centering
    \begin{subfigure}[b]{0.23\textwidth}
        \includegraphics[width=\textwidth]{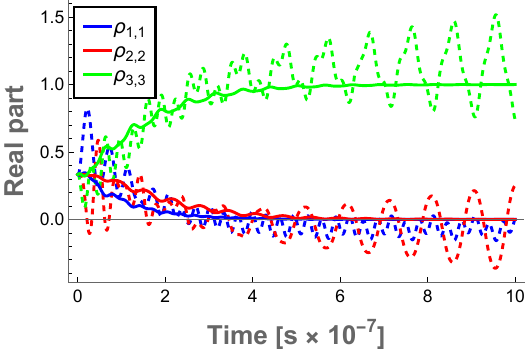}
        \caption{} \label{fig:all is wrong}
    \end{subfigure}
    \begin{subfigure}[b]{0.23\textwidth}
        \includegraphics[width=\textwidth]{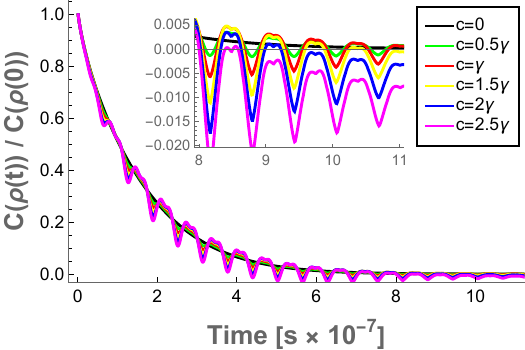}
        \caption{} \label{fig:orig c rel}
    \end{subfigure}
    \caption{Results obtained without the adjusted framework. (a) Diagonal elements of the density matrix versus time for $\omega_0 = 1 \times 10^8 \, \mathrm{s^{-1}}$, $\gamma = 0.5 \times 10^7 \, \mathrm{s^{-1}}$ and: $c=1 \times 10^7 \, \mathrm{s^{-1}}$ (solid lines) and $c=6 \times 10^7 \, \mathrm{s^{-1}}$ (dashed lines). (b) Normalized relative entropy of coherence versus time for $\omega_0 = 1 \times 10^8 \, \mathrm{s^{-1}}$, $\gamma = 0.5 \times 10^7 \, \mathrm{s^{-1}}$ and different values of $c$.}
\end{figure}

We note that the Hamiltonian \eref{eq:Seifi og hamiltonian} is non-Hermitian and therefore using \enquote{standard} framework for the density matrix evolution may lead to unphysical results. This becomes apparent when we extend the range of parameters beyond the one used by \cite{Sei22}. By numerically solving the standard Lindblad equation~\eref{eq:lindblad} for the density matrix evolution,
for value of $c$ comparable to that of $\omega_0$ and with the initial condition being the maximally coherent state
\begin{equation}
    |\psi\rangle = \frac{1}{\sqrt{3}} \left(|1\rangle + |2\rangle + |3\rangle\right) \textrm{,}
\end{equation}
we can clearly see the formalism completely breaks down. Figure \ref{fig:all is wrong} compares the evolution of the diagonal elements of the density matrix for $c=1 \times 10^7 \, \mathrm{s^{-1}}$ (solid lines) and $c=6 \times 10^7 \, \mathrm{s^{-1}}$ (dashed lines). Beyond a certain threshold, the system becomes unstable and the matrix elements, instead of settling down, start oscillating with an increasing amplitude. But even for a more conservative choice of parameters the problems emerge when inspecting the coherence of the system. Even for relatively small values of $c$, the relative entropy of coherence ventures into the negative values (cf. inset in the fig. \ref{fig:orig c rel}), which should not happen under normal circumstances, as per definition of coherence measure.

This clearly shows that the problems of the standard formalism are not only of conceptual nature -- no guarantee of the total probability conservation -- but generically lead to unphysical results.
As we will see the adjusted framework cures these problems at a minimal cost of essentially introducing an extra term in the evolution equation. For the completeness of the presentation we report here the two Hermitian operators resulting from the decomposition of the original Hamiltonian~\eref{eq:Seifi og hamiltonian}
\begin{eqnarray}
    \mathscr{H} &=& \omega_0 S_z + \frac{c}{2} \left( |2\rangle\langle 1|+ |3\rangle\langle 2| + |1\rangle \langle 3| + \mathrm{h.c.} \right) \textrm{,} \\
    \Gamma &=& \frac{i c}{2} \left( |2\rangle\langle 1|+ |3\rangle\langle 2| + |1\rangle \langle 3| - \mathrm{h.c.} \right) \textrm{,}
\end{eqnarray}
as required for the adjusted formalism. Here h.c. denotes Hermitian conjugate.

\subsection{Four states model}

The three states model is special because transitions between different states are unidirectional. Whereas directionality is crucial to model the transport phenomenon, one might wonder how stable are the predictions under introducing transitions in both ways (not necessarily with the same rate). This is the motivation between introducing the four states model.

\begin{figure}[ht]
\center
\includegraphics[width = 0.33\textwidth]{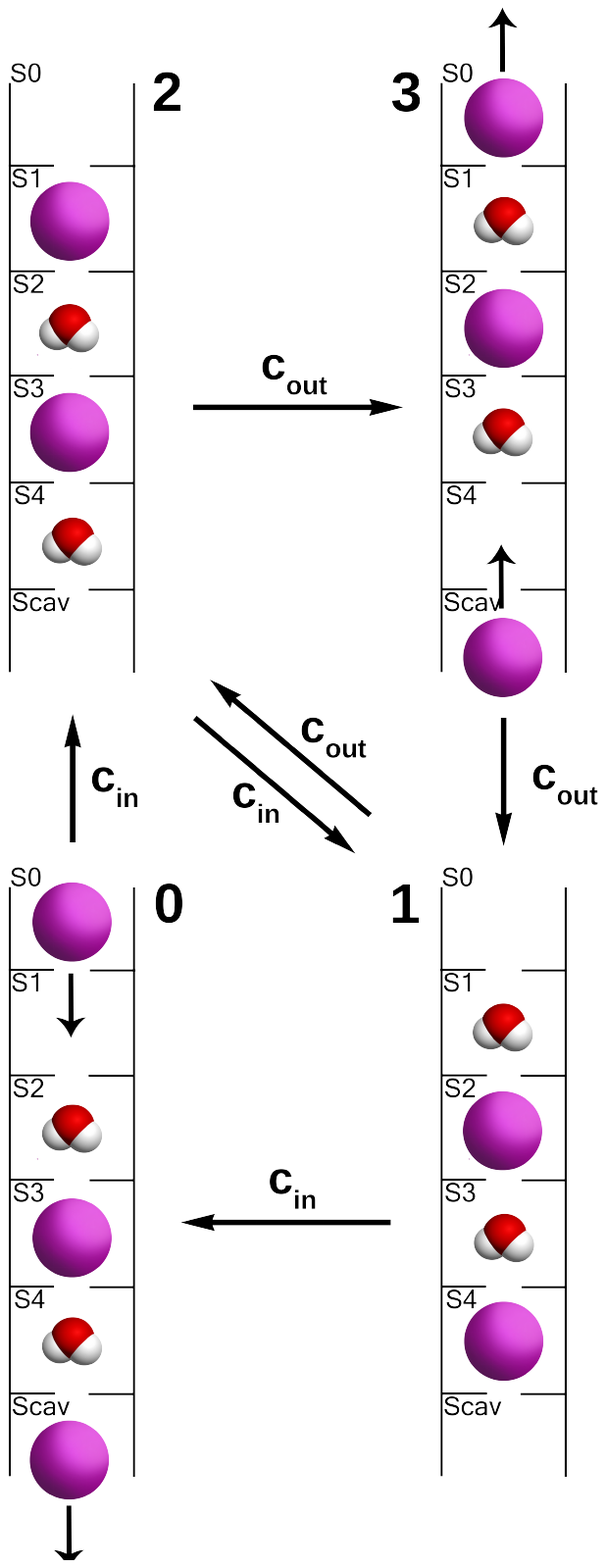}
\caption{Schematic illustration of the transition between different states in the generalized 4 state model. The arrows next to the potassium ions indicate that the ion is currently moving into the respective site. Transition rates $c_{out}$ and $c_{in}$ denote rates in the \enquote{out} pathway (ions moving in the extracellular direction) and the \enquote{in} pathway (ions moving in the intracellular direction).} \label{fig:4 states scheme}
\end{figure}

The simplest way to include transitions in both ways is to add a fourth state in which there is one ion in the S3 site of the selectivity filter, one ion entering the S1 site from the extracellular solution, and another ion leaving the channel through the cavity. We denote this state $|0\rangle$ for convenience. All permitted transitions are presented graphically in the fig.~\ref{fig:4 states scheme}. The coefficients $c_{in}$ and $c_{out}$ denote the transition rates in the pathways from the outside to the inside and vice versa. The rest of the Hamiltonian and the dissipator remain the same, but the operators are now in the spin $3/2$ representation of the $su(2)$ algebra. We will solve and analyze the four-state model in the next section in detail. For now let us confront it with the three states model on a qualitative level.

For $c_{in} = 0$ we essentially obtain the same system as in the previous sections. Still some differences can arise from the different forms of the ladder operators $S_{\pm}$ and the z-coordinate spin operator $S_z$. Without the coupling and for $c_{in} = c_{out} = c$ none of the pathways is favoured in terms of coherent transitions, but one of them may still be favoured due to the self-energy $\omega_0 S_z$. The addition of the coupling, as we already know, greatly changes the dynamics. But even though the system is now bidirectional, the exclusion of some transitions makes the Hamiltonian non-Hermitian. Therefore, we have to use the adjusted framework once again.

\section{Results}
\begin{figure}[ht]
\centering
\includegraphics[width = 0.35\textwidth]{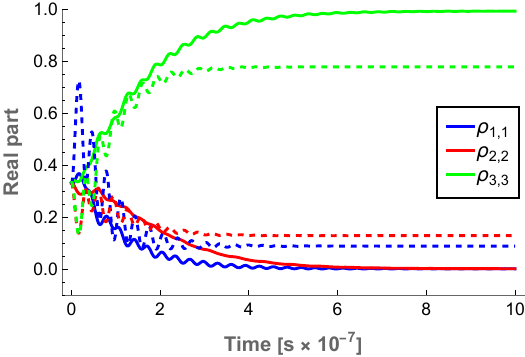}
\caption{Diagonal elements of the density matrix versus time for $\omega_0 = 1 \times 10^8 \, \mathrm{s^{-1}}$, $\gamma = 0.5 \times 10^7 \, \mathrm{s^{-1}}$ and: $c=1 \times 10^7 \, \mathrm{s^{-1}}$ (solid lines) and $c=6 \times 10^7 \, \mathrm{s^{-1}}$ (dashed lines), calculated using the adjusted framework.} \label{fig:dens evo nh}
\end{figure}

\begin{figure}[ht]
\centering
\includegraphics[width = 0.35\textwidth]{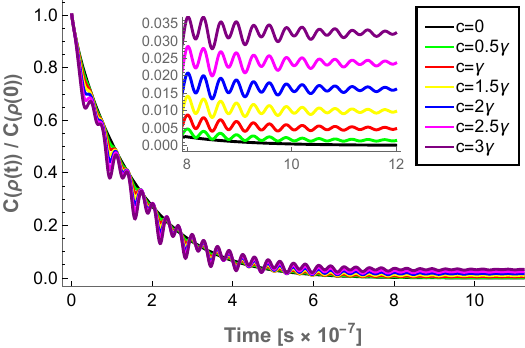}
\caption{Normalized relative entropy of coherence $C_{\mathrm{rel.ent.}}$ versus time for $\omega_0 = 1 \times 10^8 \, \mathrm{s^{-1}}$, $\gamma = 0.5 \times 10^7 \, \mathrm{s^{-1}}$ and various values of $c$.} \label{fig:crel nh}
\end{figure}

Figure \ref{fig:dens evo nh} presents the evolution of the diagonal elements of the density matrix evaluated  using the adjusted framework. For small values of $c$ the dynamics of the system remain similar to the ones obtained without the adjusted framework. For $\gamma = 0$ behaviour of the system is periodic. When incoherent coupling is included, the system tends to a quasi-stationary state (oscillations of probability become negligible), with state $|3\rangle$ most probable. For $c \sim \omega_0$ the system is now stable, but its behaviour is significantly different. The system also assumes quasi-stationary state, but the probability of the state $|3\rangle$ is considerably lower than one. Furthermore, the system becomes quasi-stationary even without the coupling, i.e. for $\gamma = 0$. We also find that the relative entropy of coherence is now always non-negative and that it remains strictly positive asymptotically (cf. fig \ref{fig:crel nh}).

Because the system is asymptotically quasi-stationary, it is sensible to ask about asymptotic behaviour of the observables as functions of the parameters of the model. In addition to the coherence measure $C$, we can investigate the conduction rate of the system, which we define as $I(t) = c \dens_{3,3}(t)$ (or $I(t) = c_{in} \dens_{3,3}(t) - c_{out} \dens_{0,0}(t)$ for the four-state model). The results for the two models and both quantities are presented in the figures \ref{fig:Sergi, crel inf, c} and \ref{fig:Sergi, i inf, c} respectively. It is no surprise that increasing the coherent hopping rate increases the asymptotic coherence, and increasing the incoherent coupling with the environment decreases it. However, the results indicate a more nuanced dynamics.

\begin{figure}[H]
    \centering
    \begin{subfigure}[b]{0.24\textwidth}
        \includegraphics[width=\textwidth]{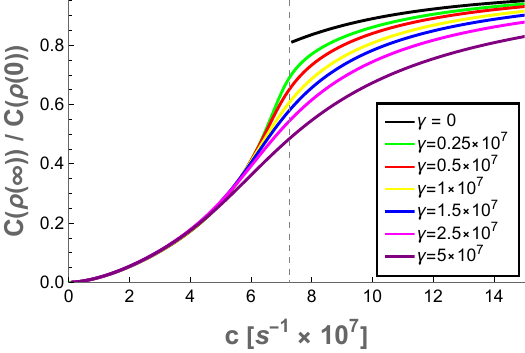}
        \caption{} \label{fig: 3st crel inf c}
    \end{subfigure}
    \begin{subfigure}[b]{0.23\textwidth}
        \includegraphics[width=\textwidth]{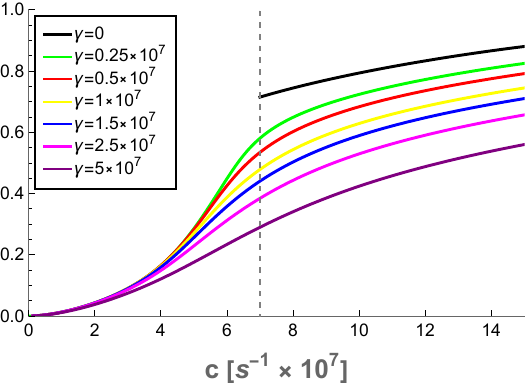}
        \caption{} \label{fig:Sergi, 4st crel ing c}
    \end{subfigure}
    \caption{Asymptotic normalized relative entropy of coherence $C(\dens (\infty))$ in the (a) three state model and (b) four state model as a function of $c$ for $\omega_0 = 1 \times 10^{8} \, \mathrm{s^{-1}}$ and various values of $\gamma$. Dashed vertical line shows the critical point for $\gamma = 0$.} \label{fig:Sergi, crel inf, c}
\end{figure}

In both models the system exhibits two regimes. The self-energy regime occurs for relatively small values of $c$. It can be associated with the situation where after a sufficient time, it is almost certain to find the system in the state $|3\rangle$, which hints very fast transitions through the states $|1\rangle$ and $|2\rangle$. Hence, asymptotic conduction increases linearly with $c$. Furthermore, the value of the coupling constant $\gamma$ (supposing it is positive) plays no role in the asymptotic behavior of the observables, since state $|3\rangle$ wins nearly all probability anyway. Yet, coupling with the environment is essential since for $\gamma = 0$ the system does not achieve any quasi-stationary state and the probabilities of different states oscillate in time with frequency and amplitude governed by $c$ and $\omega_0$. 

In the second regime, the stationary state is achieved even without the coupling, i.e. for $\gamma = 0$. There, given constant transition frequency $\omega_0$, the probabilities in the stationary state are determined by the interplay between incoherent coupling rate $\gamma$ and coherent hopping rate $c$. The larger the coherent hopping rate $c$, the more uniform the probabilities of different states tend to be. In this regime, the system dwells longer in the states $|1\rangle$ and $|2\rangle$, relatively to the state $|3\rangle$. In the limiting case, where the probabilities are equal, conduction occurs steadily through all three states. The overall speed of conduction increases with $c$. Coupling constant $\gamma$ tends to favour the state $|3\rangle$ by faster transition through the first two states. Changing $\omega_0$ changes the transition point between the two regimes.

The main difference between the three and four-state models is that in the latter one the conduction reaches plateau and eventually starts decreasing as $c$ increases. The reason is that, as the transition rate increases, transitions along the intracellular \enquote{in} pathway become more and more prominent and eventually start to contest the extracellular \enquote{out} pathway.

\begin{figure}
    \centering
    \begin{subfigure}[b]{0.24\textwidth}
        \includegraphics[width=\textwidth]{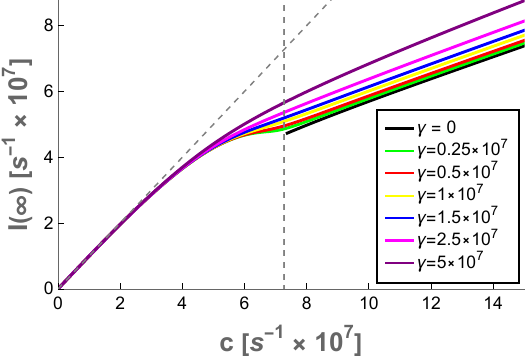}
        \caption{} \label{fig: 3st i inf c}
    \end{subfigure}
    \begin{subfigure}[b]{0.23\textwidth}
        \includegraphics[width=\textwidth]{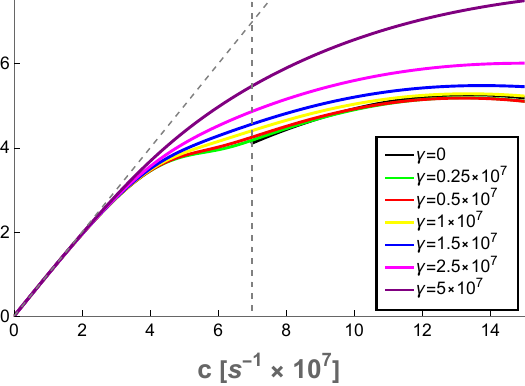}
        \caption{} \label{fig:Sergi, 4st i inf c}
    \end{subfigure}
    \caption{Asymptotic conductance $I (\infty)$ in the (a) three state model and (b) four state model as a function of $c$ for $\omega_0 = 1 \times 10^{8} \, \mathrm{s^{-1}}$ and various values of $\gamma$. Dashed vertical line shows the critical point for $\gamma = 0$, while the tilted line represents $I (\infty) = c$, i.e. linear conductance for the first regime.} \label{fig:Sergi, i inf, c}
\end{figure}

In the following section we offer an explanation for the existence of the two regimes based on the spectrum of the Hamiltonians.

\section{Hamiltonian eigenvalues}
\begin{figure}[b]
    \centering
    \begin{subfigure}[b]{0.242\textwidth}
        \includegraphics[width=\textwidth]{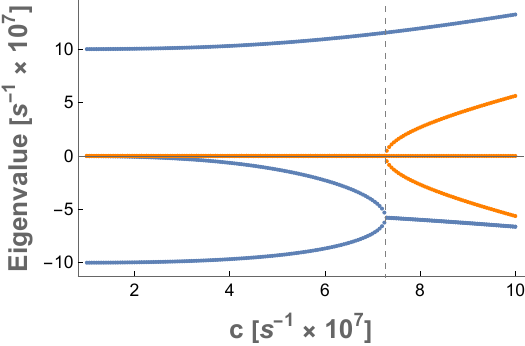}
        \caption{} \label{fig:Sergi, 3st eigenvalues}
    \end{subfigure}
    \begin{subfigure}[b]{0.23\textwidth}
        \includegraphics[width=\textwidth]{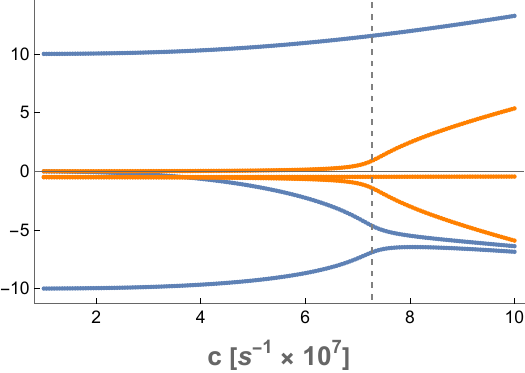}
        \caption{} \label{fig:Sergi, 3st eigenvalues with G}
    \end{subfigure}
    \caption{The real (blue) and imaginary parts (orange) of the three-state system eigenvalues as a function of $c$ for $\omega_0 = 1 \times 10^8 \, \mathrm{s^{-1}}$ and (a) $\gamma = 0$, (b) $\gamma = 0.5 \times 10^{7} \, \mathrm{s^{-1}}$. The critical point for $\gamma = 0$ is shown by the dashed line.} \label{fig:Sergi eigenvalues}
\end{figure}

Consider the Hamiltonian \eref{eq:Seifi og hamiltonian} itself without the coupling to the environment described by the Lindblad operator. Despite its non-hermiticity, we can still solve for its spectrum. In fact, due to the relatively easy form of the Hamiltonian in the three-state model, we can obtain an analytical solution for its eigenvalues
\begin{eqnarray}
    \varepsilon_1 &&= \frac{2 \cdot 3^{1/3} \cdot \omega_0^2 + 2^{1/3} \cdot \alpha^{2/3}}{6^{2/3} \cdot \alpha^{1/3}} \nonumber \textrm{,} \\
    %\varepsilon_2 &&= \frac{-(3^{1/3} + 3^{5/6} i) \omega_0^2 + (-1)^{1/3} \cdot (-2)^{1/3} \cdot \alpha^{2/3}}{6^{2/3} \cdot \alpha^{1/3}} \textrm{,} \\
    \varepsilon_2 &&= \frac{-(3^{1/3} + 3^{5/6} i) \omega_0^2 + 2^{1/3} \cdot \alpha^{2/3}}{6^{2/3} \cdot \alpha^{1/3}} \textrm{,} \\
    \varepsilon_3 &&= - \frac{(3^{1/3}-3^{5/6} i) \omega_0^2 + (-2)^{1/3} \cdot \alpha^{2/3}}{6^{2/3} \cdot \alpha^{1/3}} \nonumber \textrm{,}
\end{eqnarray}
where
\begin{equation}
    \alpha = 9c^3 + \sqrt{81 c^6 - 12 \omega_0^6} \textrm{.}
\end{equation}
The spectrum is shown in the fig. \ref{fig:Sergi, 3st eigenvalues} for $\omega_0 = 1 \times 10^{8} \, \mathrm{s^{-1}}$. The blue color denotes the real parts of the eigenvalues, and the orange shows the imaginary parts. At some point two of the eigenvalues coalesce, and their imaginary parts become non-zero. Such points are called exceptional and they play an important role in, among others, $\mathcal{PT}$-symmetric non-Hermitian quantum mechanics, photonics and optics~\cite{Heiss12, Miri19}. This transition happens at the point, where $\alpha$ becomes purely real, i.e. at $c = (12/81)^{1/6} \omega_0 \approx 0.72742 \omega_0$. A similar thing happens in the four-state model, with two of the eigenvalues remaining real beyond the exceptional point (not shown).

Turning on the coupling to the environment has an impact on the dynamics of the system. Nevertheless, the two regimes in this case should be the result of a similar phenomenon. To this end, we need to incorporate to the Hamiltonian, at least partially, the Lindblad term. This can be achieved by the technique known from non-Hermitian quantum mechanics where it is a common practice to partially include the Markovian part into the effective Hamiltonian. The effective Hamiltonian becomes then non-Hermitian (if it has not been already). The prescription is~\cite{Roc22},
\begin{equation}
    H_{\mathrm{eff}} = H - \frac{i}{2} \sum_k \Gamma_k L^{\dagger}_k L_k \textrm{,}
\end{equation}
where $L_k$ are the Lindblad operators as in \eref{eq:lindblad}. Then the density matrix evolves according to the original Hamiltonian with the addition of the so-called quantum jumps $\Gamma_k L_k \dens L^{\dagger}_k$ \cite{Roc22}. We introduce such effective Hamiltonian in the three-state model
\begin{equation}
    H_{\mathrm{eff}} = H - \frac{i}{2} \gamma S_{+} S_{-}
\end{equation}
and plot its eigenvalues for $\gamma = 0.5 \times 10^7 \, \mathrm{s^{-1}}$. The result is shown in the fig. \ref{fig:Sergi, 3st eigenvalues with G}. Although the imaginary parts of the eigenvalues are now shifted and therefore non-zero, we can still identify two distinct regimes. In the first one, the imaginary parts are orders of magnitude smaller than the real parts and vary slowly. In the second one two of them start to diverge quite rapidly. 

As we will now show the eigenvalue structure determines the dynamics and, more specifically, is responsible for distinct asymptotic regimes observed in fig.~\ref{fig:Sergi, crel inf, c} and~\ref{fig:Sergi, i inf, c}. 
Let us start with the evolution of the density matrix for non-Hermitian closed system as given by eq.~\eref{eq:nonhermitian time evolution}. Simple manipulations yields
\begin{equation}
    \frac{\partial \rho}{\partial t} = - i \left(H_{\rm eff} \rho - \rho H_{\rm eff}^{\dagger} \right) + i \rho\, \Tr\left(H_{\rm eff} \rho - \rho H_{\rm eff}^{\dagger} \right) \textrm{,}
\end{equation}
with the second term enforcing the normalization $\Tr \rho = 1$. At large times, the dynamics is dominated by the eigenstates with the largest imaginary eigenvalue. This can be seen by writing the first term in the energy eigenbasis where it becomes $-i \left( E_a  - E_b^*\right)\rho_{ab}$. Here, we introduced eigenstates $|a\rangle$ such that $H_{\rm eff}|a\rangle = E_a |a\rangle$ with $E_a$ potentially complex and $\rho_{ab} = \langle a |\rho |b\rangle$. Therefore, at large times system evolves towards maximizing the imaginary part of $E_a - E_b^*$. However
\begin{equation}
    \max_{a, b} \, {\rm im} (E_a - E_b^*) = 2\max_{a} \, {\rm im} E_a \textrm{,}
\end{equation}
which shows that this is indeed achieved by choosing an eigenstate with the largest imaginary eigenvalue. The eigenspectrum, as shown in Fig.~\ref{fig:Sergi eigenvalues}, reveals that the largest imaginary eigenvalue undergoes a transition from mainly flat and small to quickly increasing defining the two regimes visible in the relative entropy of coherence and conductance. 

\section{Discussion}

In this work, we have studied the \enquote{soft knock-on} model of transport through ion channels in the quantum-mechanical setting. We have argued that the non-hermiticity of the Hamiltonian requires modifying the usual Lindblad formalism. We have used the "hybrid" formalism of~\cite{ZloSer13} which applies to non-hermitian systems coupled to an environment. The results revealed two regimes of the model visible at long times. At small transition rates, $c/\omega_0 \ll 1$, the stationary state is indifferent to the coupling with the environment. In this regime conductance increases linearly with the transition rate $c$. On the other hand, for $c > \omega_0$ the coherence and conductance of the stationary state depend on the environmental noise. The noise increases the transport, but at the same time has a detrimental effect on the coherence. Still, the coherence, as witnessed by the relative entropy, remains large enough for the transport phenomena to be in a quantum regime. 

Thus, the results show there is no contradiction between the high level of coherence and high conductance. Depending on the values of the parameters of the model, the transport properties of ion channels are determined either by the internal conductance mechanism or by the interplay between it and the external noise. Resolving which way the transport occurs in actual ion channels would require a more precise estimation of the parameters from the experiments. We have also offered an explanation of the existence of the two regimes based on the spectrum of the effective Hamiltonian. As a by-product of using the adapted formalism for the evolution of the density matrix, we have solved the problem of unphysical results as seen in~\cite{Sei22}.

From a wider perspective, we have demonstrated that the "hybrid" formalism is an effective tool to describe non-Hermitian open systems, which are widespread in biophysical context.  

Finally, it would be also interesting to perform further analysis for the competing \enquote{hard knock-on} model, beyond the already existing work \cite{Vaz10, Mar18, Mar21}, which might help discriminate between the two based on experimental evidence.

{\bf Acknowledgments:} We thank Bert de Groot for useful discussions.

\section*{References}

%\begin{thebibliography}
%\end{thebibliography}
\bibliographystyle{unsrt}
\bibliography{Bibliography}

\end{document}